# Intrinsic Half-metallicity in in Edge-fluorinated Armchair Boron Nitride Nanoribbons (arXiv:1501.01473v4)

## Published/Present Title: - Possibility of Spin-polarized Transport in Edge-fluorinated Armchair Boron Nitride Nanoribbons

*Hari Mohan Rai[1], Shailendra K. Saxena[1], Vikash Mishra[1], Ravikiran Late[1], Rajesh Kumar[1], Pankaj R. Sagdeo[1],*

[1]Material Research Laboratory, Discipline of Physics, Indian Institute of Technology Indore, Simrol, Indore (M.P.) – 452020, India

*Neeraj K. Jaiswal[2]*

[2]Discipline of Physics, PDPM- Indian Institute of Information Technology, Design and Manufacturing, Jabalpur – 482005, India

*Pankaj Srivastava[*3]*

[3]Computational Nanoscience and Technology Lab. (CNTL), ABV- Indian Institute of Information Technology and Management, Gwalior – 474015, India

\* Corresponding author:  pankajs@iiitm.ac.in

ABSRTACT

We predict the possibility of spin based electronic transport in edge fluorinated armchair boron nitride nanoribbons (ABNNRs). The structural stability, electronic and magnetic properties of these edge fluorinated ABNNRs have been systematically analyzed by means of first-principles calculations within the local spin-density approximation (LSDA). Regardless of their width, ABNNRs with F-passivation at only edge-B atoms are found to be thermodynamically stable and half-metallic in nature. The spin polarized states are found to be ~0.4 eV more stable than that of spin compensated states. Further, upto 100% spin polarization is expected in ABNNRs with F-passivation at only edge-B atoms as indicated by giant splitting of spin states which is observed in corresponding band structures, DOS and transmission spectrum. The existence of half-metallicity is attributed to the localization of electronic charge at unpassivated edge-N atoms as revealed from the analysis of Bloch states and projected density of states (PDOS). Importantly, present stability analysis suggests that the possibility of experimental realization of spin polarized transport in ABNNRs is more promising via F-passivation of ribbon edges than that of H-passivation. The observed half-metallic nature and large difference in the energies (~0.4 eV) of spin polarized and spin compensated states projects these half-metallic ABNNRs as potential candidate for inorganic spintronic applications.

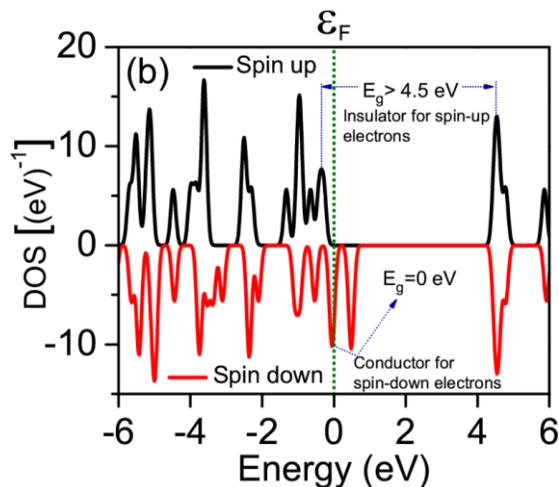

Calculated DOS for edge-fluorinated ABNNRs; featuring Half-metallicity.

**TOC**

# 1. INTRODUCTION

The discovery of single atom thick sheet of hexagonal boron nitride (h-BN) has invigorated theoretical as well as experimental research activities associated with h-BN nanomaterials[1–4]This quasi-two-dimensional (2D) h-BN sheet has large energy band gap (~5eV) which ensures its stability at elevated temperatures and thus makes it a promising candidate for high temperature semiconductor devices. Some interesting features like capability of dielectric screening,[5] strong oxidation resistance[6] and tunable (achieved via creating triangular defects)[7] magnetic and electronic properties are recently reported in atomically thin BN nanosheets. Quasi-1-D thin strips carved out of h-BN sheet also exhibit excellent structural stability and distinct electronic properties as compared to parent h-BN sheet due to the quantum confinement effects [4]. These strips are formally known as boron nitride nanoribbons (BNNRs). Chen. et al. [8] have already synthesized hollow BNNRs through evaporation using B–N–O–Fe as precursor and ZnS nanoribbons as templates. Moreover, BNNRs have also been produced experimentally via in-situ unzipping of boron nitride nanotubes (BNNTs) [9]. Similar to their organic counterpart i.e. Graphene Nanoribbons (GNRs), BNNRs can also be fabricated from a single h-BN layer via lithographic patterning [10]. According to the definite shape of edges BNNRs are mainly of two type i.e. zigzag boron nitride nanoribbons (ZBNNRs) and armchair boron nitride nanoribbons (ABNNRs). BNNRs exhibit fascinating and outstanding electronic properties similar to that of GNRs, which are crucial for the application in advance devices [11,12]. For example; Cl adsorption [13] and $PH_3$-passivation of ribbon edges [14] significantly modifies electronic state/band gap of ABNNRs which can be useful in gas sensor technology for sensing Cl and $PH_3$ gas molecules. The electronic band gap ($E_g$) of ABNNRs passivated by H, oscillates periodically as a function of their width [15] whereas the band gap of ZBNNRs decreases monotonically, with increasing ribbon, width when both the edges are H-passivated [16,17]. Owing to high efficiency and low power consumption, spintronic devices are attracting considerable attention nowadays. Being the key property for spin transport based electronics, Half-metallicity, has been vastly predicted and studied in different nanostructures [18–20] along with many other compounds like alloys (including ternary and quaternary Heusler alloys) [21–23], double layer perovskites [24,25], transition metal pnictides, and chalcogenides in zinc blende phase [26–28]. Recently, the existence of a large spin polarization of $(93^{+8}_{-11})\%$ has been directly evidenced at room temperature in an epitaxial thin film of $Co_2MnSi$ [29]. First principle calculations have already shown that half-metallicity may be realized also in BNNRs either by chemically functionalizing zigzag edges with different groups such as H and F or with the application of an external in-plane electric field [30–32]. In previous study on ZBNNRs [16], it is predicted that one-edge (only B-edge) H-passivation makes the ribbons 'semi-metallic' unlike the 'half-metallic' as predicted by Zheng et al. [30] but in the perspective of spin polarized transport, both of these results might be consistent with each other. Moreover, half-metallicity has also been perceived theoretically in ZBNNRs via percentage hydrogenation [33], hybridization (through carbon in $BC_2N$ nanoribbons) [34], and also through structural modifications (e.g., by making stirrup, boat, twist-boat etc.) [35]. Conversely, in case of ABNNRs, half-metallicity has been predicted in hybrid armchair BCN- nanoribbons [36] and also in H-passivated (edge-hydrogenated) ABNNRs [37] but the experimental realization of these edge hydrogenated ABNNRs is energetically not favorable as they are thermodynamically less stable [37], However, from the perspective of energetically stable spin-polarization, ABNNRs are almost unexplored in comparison with zigzag counterparts. By keeping in view, that ABNNRs can also be potential spintronics elements, this article presents an attempt - to predict the possibility of spin polarized transport in ABNNRs via edge fluorination and to analyze it in detail using DFT calculations.

Here, we predict the possibility of spin polarized transport in ABNNRs when only edge-B atoms are passivated with F atoms. In addition, this article presents a systematic analysis of structural stability, electronic and magnetic properties for fully and partially edge fluorinated ABNNRs with different widths. Present analysis reveals that edge fluorination transforms bare ABNNRs into ferromagnetic/antiferromagnetic half-metals as energetically most favorable structures unlike edge-hydrogenated counterparts [37,38].

## 2. COMPUTATIONAL DETAILS

The first-principles calculations were performed with Atomistix Tool Kit-Virtual NanoLab (ATK-VNL) [39,40] under the framework of density functional theory (DFT). The exchange correlation energy was approximated by local spin density approximation (LSDA) as proposed by Perdew and Zunger [41]. The reason for selecting LDA for present DFT calculations is that the generalized gradient approximation underestimates the surface–impurity interactions [42]. ATK employs an effective potential $V^{eff}[n] = V^{H}[n] + V^{xc}[n] + V^{ext}$ for calculations. Here, first two terms exhibit the potential due to interactions with the other electrons (represented in terms of electron density); the first term, $V^H[n]$, is the classical electrostatic potential (also known as Hartree potential) due to the mean-field electrostatic interaction, and the second term, $V^{xc}[n]$, is the exchange-correlation potential caused by the quantum mechanical nature of the electrons. The third term $V^{ext}$ represents external potential which is the sum of ions potentials (represented by norm-conserving pseudo-potentials) and electrostatic interaction of system with an external electrostatic field. For present calculations ABNNRs were modeled with periodic boundary conditions along *z*-axis, whereas the other two dimensions were confined. Norm-conserving pseudo-potentials with an energy cut off of 100 Ry, was selected for the expansion of plane waves. We implemented double $\zeta$ plus polarized basis set for all the calculations. The *k*-point sampling was selected to be 1×1×100. In order to avoid artificial inter-ribbon interactions, ribbons were separated using a cell padding vacuum region of 10 Å. All the atoms, in the considered geometries, are relaxed and optimization of atomic positions and lattice parameters has been continued until the forces on each constituent atom reduced upto 0.05 eV/Å. We represent the ribbon-width by a width parameter $N_a$, defined as the number of B or N atoms [37] along the ribbon width as depicted in figure 1, therefore, ABNNR with n B or N atoms across the ribbon is named as *n*-ABNNR.

## 3. RESULTS AND DISCUSSION

For present calculations we have considered ABNNRs with – (i) fully F-passivated edges (ABNNR$_{FBN}$) and (ii) partial/half F-passivated edges. Furthermore, the ABNNRs belonging to latter category are divided into two subgroups i.e., ABNNR$_{FB}$ and ABNNR$_{FN}$ depending upon whether only B or only N edge atoms are passivated by F-atoms respectively. The electronic and magnetic properties of bare ABNNRs have also been investigated and these ribbons were found to be nonmagnetic semiconductors as also predicted by Ding *et al.* [38]. In order to take size effects into considerations, we investigate ABNNRs having widths $N_a$ = 6 to 10. Since the findings are qualitatively similar for symmetric ribbons (Odd- $N_a$), we displayed the figures only for ABNNRs with $N_a$ = 9. Figure 1 schematically represents optimized geometry of ABNNR in all considered ribbon configurations with $N_a$ = 9 as representative case. The convention of super cell, used for simulation, is also depicted in the same figure for various ribbon structures.

## 3.1 Stability analysis

First of all, in order to examine the relative stability of considered ribbon configurations, cohesive energy ($E_c$) has been calculated by using the following relation;

$$E_C = \frac{1}{(p+q+r)}\left[E_{tot}^{ABNNR} - \left(pE_{atom}^{B} + qE_{atom}^{N} + rE_{atom}^{F}\right)\right] \quad (1)$$

where $E_{tot}^{ABNNR}$, $E_{atom}^{B}$, $E_{atom}^{N}$ and $E_{atom}^{F}$ are the total energies of ABNNR, free B atom, N atom and F atom, respectively; p, q and r are the number of B, N and F atoms respectively in the super cell. The cohesive energies calculated for considered ribbon structures are summarized in Table I. It is noticeable that lower values of $E_c$ favor higher stability. The perusal of Table I indicates that irrespective of their width ABNNR$_{FB}$ are energetically most stable ribbon structures amongst all considered configurations and the minimum difference in $E_c$ is found to be 330 meV which is observed for 10-ABNNR$_{FB}$ and 10-ABNNR$_{FN}$ structures. This difference is large enough for a room temperature operation. Further, the energy difference is more pronounced at narrow widths whereas wider ribbons are found to be relatively more stable [Table 1]. This indicates that amount of stress in the nanoribbons, which causes structural instability, is inversely proportional to the ribbon width. Present stability analysis predicts that amongst all investigated edge fluorinated ABNNRs; the possibility of experimental realization is highest for ABNNR$_{FB}$ structures as they have lowest $E_c$. It is noticeable that we observed a mild reconstruction at the outer edges of ribbons when only B/N edge-atoms are saturated with F (i.e., ABNNR$_{FB}$ and ABNNR$_{FN}$). Owing to this edge reconstruction the bond length of the outer edge B-N bond is slightly reduced (1.37 Å) as compared to the normal bond length (1.38Å) which provides stability to these ribbon structures in a similar way to that of edge reconstructed GNRs where edge reconstruction was responsible for the edge stability of bare GNRs [43]. In addition, it should be noted as an extra information that edge-fluorinated ZBNNRs are energetically more favorable than that of edge-fluorinated ABNNRs as corresponding cohesive and binding energies are tabulated in supplementary information (Table S1 and Table S2). Moreover, it has been already reported that H-passivation (hydrogenation) of only B edge atoms (ABNNR$_{HB}$) makes armchair ribbons half-metallic [37] but higher value of $E_c$ restricts their experimental realization because in case of edge hydrogenation, armchair ribbons with H-passivation at only N edge atoms (ABNNR$_{HN}$) were found relatively more stable (as the value of corresponding $E_c$ was found to be lowest) than that of ABNNR$_{HB}$ [37]. Importantly, in the present case of edge-fluorination (F-passivation); armchair ribbons with F-passivation at only B edge-atoms are found to be half-metallic and most stable (ABNNR$_{FB}$) amongst all presently investigated ribbon configurations. In addition to this $E_c$ analysis, binding energy ($E_b$) per passivator–atom has also been calculated to further verify the stability (i.e., the probability of experimental realization of passivation) of bonding between Passivator atoms and host edge B/N atoms. For this purpose, both, F and H-atoms are used separately as edge-passivators in different ribbon configurations and $E_b$ per F (or H)-atom is calculated by using following formula-

$$E_B = \frac{1}{x}\left[E_{tot}^{Configuration} - E_{tot}^{Bare} - xE_{tot}^{F(orH)}\right]$$

Where $E_{tot}^{Configuration}$, $E_{tot}^{Bare}$ and $E_{tot}^{F(orH)}$ are the total energies of F (or H)-passivated ABNNR, bare ABNNR and isolated F (or H) atom respectively; $x$ is the number of F (or H) atoms present in the ribbon structures under investigation. The calculated values of $E_b$ for all three considered F (or H)-passivated armchair ribbons, with varying ribbon width, are summarized in Table II and the values in brackets are corresponding to H-passivation. It is to be noted that lower $E_b$ represents higher stability. The perusal of Table II indicates that irrespective of their width ABNNR$_{FB}$ structures are energetically most stable with lowest $E_b$ amongst all considered F-passivated ribbon configurations whereas in case of H-passivation; ABNNR$_{HB}$ are found to be less stable relative to their remaining H-passivated counterparts (i.e., ABNNR$_{HN}$ and ABNNR$_{HBN}$). Though ABNNR$_{HB}$ are also half-metallic [37], however, this comparative analysis between F and H-passivated ABNNRs reveals that the possibility of achieving spin polarized transport experimentally is more favorable in edge-fluorinated ABNNRs (ABNNR$_{FB}$) than that of edge-hydrogenated ones (ABNNR$_{HB}$). Thus, being energetically most favorable ribbon structures amongst all presently investigated edge-fluorinated ABNNRs, the results for all ABNNR$_{FB}$ structures are addressed with greater detail (relative to other counterparts) in rest of the text.

3.2 Electronic and transport properties

The calculated electronic band structure and density of states (DOS) along with the Transmission spectrum (TS) [11,16] for 9-ABNNR$_{FB}$ (as a representative case) are illustrated in figures 2 (a), 2 (b) and 2 (c) respectively. A substantial difference in the form of large splitting of spin states can be clearly observed in the band structures as well as in DOS profile and TS. The spin-down states are metallic as red bands [figure 2 (a)] are crossing the Fermi level near the zone center and zone boundary whereas owing to the existence of a large band gap of about 4.75 eV spin-up (black) states exhibit insulating behavior. A small peak in transmission coefficient (TC=3) corresponding to spin down transmission channels is observed [figure 2 (c)] across the Fermi level which points towards the presence of conducting channels for spin down electrons whereas an energy window of about 4.5 eV corresponding to spin up transmission states (i.e., separation between spin up transmission states across Fermi level) indicates towards the unavailability of conducting paths for spin up electrons. In addition, the absence of mirror symmetry [16] between up and down spin DOS/transmission states [figure 2 (b) and 2 (c)] is correlated with the splitting of spin states as observed in band structure $ABNNR_{FB}$ structures [figure 2 (a)]. Thus, we predict, on the basis of electronic band structure [figure 2 (a)], that the charge transport across the Fermi level through all $ABNNR_{FB}$ structures will be completely dominated by the spin-down electrons as it is evident also from DOS profile and TS [figures 2 (b) and 2 (c) ] in terms of – (i) crossing of Fermi level by a peak corresponding to spin-down states and (ii) large separation (> 4 eV) between spin up DOS/ transmission states across Fermi level (i.e. spin up conduction band minima (CBM) and spin up valence band maxima (VBM) . Therefore, it is expected that the current flow though such systems should be entirely spin polarized. Moreover, a half-metal gap (i.e. the gap between top most occupied spin-up band (α' for present case) and Fermi level [30]) of about 0.21 eV has been witnessed which is high enough for a room temperature device operation as it is sufficiently large than that of the room temperature thermal energy (26 meV) and it is also comparable to the earlier reported values for half-metallic ZBNNRs and GNRs [30,44].

Figure 3 (a), 3(b) and 3 (c) respectively display the electronic band structures along with the corresponding DOS and TS for $ABNNR_{FBN}$ configurations. It is obvious from figure 3 (a) that $ABNNR_{FBN}$ are insulating ($E_g$>4.2 eV) in nature with degenerate electronic bands. The DOS profile also support this insulating behavior in terms of a large separation between the peaks of corresponding spins across Fermi level. The TS, as depicted in figure 3(c), exhibits complete absence of conduction channels around Fermi level as the value of TC is zero across green dotted line and the large separation of transmission states of corresponding spins across Fermi level points towards the existence of a wide band gap in $ABNNR_{FBN}$ structures. The width dependence of band gap for $ABNNR_{FBN}$ structures is illustrated in the inset of figure 3 (a). An oscillating behavior of band gap with increasing ribbon width is observed for ABNNR$_{FBN}$ structures which is a characteristic signature of ABNNRs [15]. Moreover, the mirror symmetry in the peaks of opposite spin as appearing in the DOS and transmission states [figure 3(b) and 3(c)], is due to preserved spin degeneracy. In contrary, ABNNR$_{FN}$ shows breaking of spin degeneracy asfdepicted in the corresponding band structure [figure 4(a)] which is further supported by corresponding DOS profile and TS [figure 4(b) and 4(c)]. All ABNNR$_{FN}$ structures, regardless of their width, exhibit a spin dependent wide band gap; an example of 9-ABNNR$_{FN}$ is shown in figure 4 as a representative case.

Now, with this brief discussion on electronic and transport properties, we again come back to half-metallic and energetically most favorable ribbons (i.e. ABNNR$_{FB}$) amongst all presently investigated configurations. In order to understand the origin of half- metallicity in ABNNR$_{FB}$ structures, we analyzed that how the passivation of edge atoms affects the band structure and DOS. The passivation of edge N atoms with F mainly shifts α band away from the Fermi level while change in the position of β band depends also upon passivation of edge-B atoms (figures 2-4). Projected DOS (PDOS) analysis reveals that both of these bands are predominantly composed of 2$p$ orbital of unpassivated edge N atoms. Moreover, we have estimated the bare on site Coulombic repulsion [31,45] as a difference between the first order ionization energy and electron affinity [31,46] for each element individually. The estimated values are 14.01 eV for F, 8.02 eV for B and 14.79 eV for N. This large difference ($\geq$ 5.99 eV) between N/F and B atoms inhibits the charge transfer through edge-B atoms/B-F bond; therefore, the electrons are localized only in 2$p$ orbital of N atoms. This mechanism for origin of half-metallicity is consistent with PDOS analysis and previous reports [31,47] on BNNRs.

3.3 Partial charge analysis (Bloch states analysis)

In order to confirm the localization of charges (which is associated with the origin of presently observed half-metallic nature of ABNNR$_{FB}$) in unpassivated edge-N atoms of ABNNR$_{FB}$, Bloch states have been calculated and the same has been shown in figure 5. The structure of 9-ABNNR$_{FB}$ [figure 1 (a)] consists of 27 B, 27 N and 6 F atoms. A single B (N) atom contributes three (five) valence electrons as its electronic configuration is $2s^22p^1$ ($2s^22p^3$) whereas a single F atom has seven valence electrons - $2s^22p^5$. Thus there are total 258 valence electrons in the system. Moreover, each band is doubly degenerated (considering spin) and hence there will be a total of 129 valence bands. Therefore, the band index which we used in calculations for VBM is 128 (α band) and 129 (β band) for CBM as indexing starts from zero. For convenience, the spin-down band existing just below (above) α (β) band [figure 2 (a)], is designated as (α-1) [(β+1)] band. The band index of spin-down (α-1) and (β+1) bands are 127 and 130 respectively whereas the band index for spin up α' and β' bands are 126 and 131 respectively. The presented Bloch states calculations are performed for k-points value of (0.0, 0.0, 0.5). In order to keep the size of figures in presentable format, the contour and isosurface of 9-ABNNR$_{FB}$ for α, β, (α-1) and (β+1) bands are shown only with one third of the unit cell as depicted in figure 5(a) and 5(b) respectively. For a clear view, the results of β band are also presented with complete unit cell which is used for calculations. It is clear that the Bloch states (red and dark yellow orbitals) corresponding to (α-1) and (β+1) bands are completely distinguishable as no mixing of red and yellow orbitals is present in corresponding isosurface [figure 5 (b)]. Similar results are obtained for spin up α' and β' bands of ABNNR$_{FB}$ structures (not shown). These results are also consistent for ABNNR$_{FN}$ and ABNNR$_{FBN}$ (not shown) as none of the bands crosses Fermi level in both of these ribbon configurations. Moreover, the contour and isosurface (for α and β bands particularly) are almost identical (considering charge localization) as both of these bands are correspond to the same spin state (downspin) present across the Fermi level. It can be observed from contour and isosurface plots that the unpaired electrons are mainly localized at the unpassivated edge-N atoms. This charge localization suggests that dangling bond of the edge N-atoms is mainly responsible for observed half-metallic behavior of ABNNR$_{FB}$ structures. In addition, the interchanged colors of Bloch states for α and β bands, at top ribbon edge (as shown by solid black arrows), probably points towards a distinct separation of the corresponding bands (α and β) with Fermi level as it is also similar to the case of interchanged colors of Bloch states (as indicated by dotted black arrows) corresponding to (α-1) and (β+1) bands which are well separated with Fermi level [figure 2(a)]. Conversely, colors, at the bottom ribbon edge (as displayed with solid violet arrows), are same for α and β bands which points towards the encroachment of α band (β band) into CB (VB). This suggests that the metallic behavior for spin down states is mainly associated with the localized charge present at unpassivated edge N-atoms due to the unpaired electrons. Moreover, it can be seen that a little charge localization is also appearing around F and near edge N atoms but it has no contribution across the Fermi level as confirmed via PDOS analysis. In addition, the distance between any two adjacent edge N atoms (~2.43 Å) is large enough to ensure that no intense edge reconstruction is taking place (note-occurrence of which could have resulted into two or more dangling bonds per edge-N atom) due to minimum separation between any two adjacent edge N atoms and consequently only one dangling bond is present per edge-N atom. Thus, the present partial charge analysis (Bloch states analysis) suggests that the half-metallic spin polarization in ABNNR$_{FB}$ structures is entirely attributed to the localization of unpaired electrons at unpassivated edge-N atoms.

Further, we calculated electron density for ABNNR$_{FB}$ and ABNNR$_{FN}$ structures. The real space distribution of calculated electron density for ABNNR$_{FB}$ and ABNNR$_{FN}$ structures is depicted in figure 6 and figure S1 (supplementary information) respectively. It has been observed from these figures that for ABNNR$_{FB}$, electron densities corresponding to spin up electrons are almost identical for all N atoms present in the ribbon whereas for spin down electrons the density at unsaturated edge N-atoms is different (as indicated by arrows) than that of all other N atoms present in the same ribbon with almost identical electron densities. This suggests that in case of ABNNR$_{FB}$, spin down electrons present at unsaturated edge N-atoms are responsible for conduction as also confirmed through corresponding band structure, DOS and TS. It is noticeable that in case of ABNNR$_{FN}$, for all atoms present in the ribbon, the electron density is found to be almost identical irrespective of electron spins (figure S1 of supplementary information) which supports to the separation of electronic states across Fermi level as also validated though corresponding band structure, TS and DOS. Present analysis of real space distribution of spin dependent electron densities suggest that the possibility of half-metallic spin polarization in ABNNR$_{FB}$ structures is completely attributed to the localization of unpaired electrons at unsaturated edge-N atoms.

## 3.4 Magnetic Properties

Since, the sable magnetic state of the material is important form the perspective of spin polarized transport; therefore, we have analyzed the effect of different kinds of edge F-passivation on magnetic properties of ABNNRs. The magnetic properties like magnetic moment for all presently studied ABNNRs have been determined through Mullikan population calculations [11] and the calculated magnetic moment M ($\mu_B$) along with the energy difference $\Delta E$ (eV) between spin polarized ($E_{LSDA}$) and spin compensated states ($E_{LDA}$) for the most stable edge fluorinated ribbon configuration (ABNNR$_{FB}$) are presented in Table III. A noteworthy net magnetic moment of about 2.03 µ$_B$ is observed for all ABNNR$_{FB}$ structures. In addition, a net magnetic moment of about 2.02 µ$_B$ is also perceived in ABNNR$_{FN}$. On the contrary, any of the ABNNR$_{FBN}$ structures does not exhibit any magnetism. It is worth noticing here that the latter magnetic moment probably have little practical significance as it is observed in ABNNR$_{FN}$ structures which are less stable than that of ABNNR$_{FB}$ structures as confirmed through E$_C$ and E$_b$ analysis. Present non-$d$ orbital type magnetism observed in ABNNR$_{FB}$ structures is accounted due to the magnetic moment of unpaired electrons localized in unpassivated edge-N atoms. This is consistent with giant splitting of spin states [figure 2] and localization of electrons as discussed in Bloch state analysis. It is clear from Table III that the spin polarized state is ~0.4 eV ($\Delta E=|E_{LSDA}-E_{LDA}|$) more stable with respect to the spin compensated state ($E_{LDA}$). This difference is large enough for room temperature device applications and it is also comparable to the values (0.17 eV, 48 meV and 3.5 meV) reported previously [30,33,35] for ZBNNRs. This significant energy difference offers a strong possibility for presently studied half-metallic ABNNR$_{FB}$ structures to be used as potential candidates for inorganic spintronic device applications. Moreover, the coupling between electrons of inter-edge unpassivated N atoms may be either ferromagnetic or antiferromagnetic, as the observed energies are very close to each other [35]. Ultimately, we reveal that electronic and magnetic properties of ABNNRs critically depend on the type of edge fluorination, as summarized in Table IV.

## 4. CONCLUSIONS

Conclusively, the effect of edge fluorination on the electronic and magnetic properties of ABNNRs has been investigated systematically using ab-initio calculations. ABNNR$_{FB}$ are found to be half-metallic and energetically most favorable, with a half-metal gap of 0.21 eV. The spin polarized states are found to be more stable by an energy of ~0.4 eV as compared to spin compensated states (E$_{LDA}$) which projects these half-metallic ribbons as promising building blocks for inorganic spin based electronic (spintronic) devices. Importantly, present stability analysis suggests that the possibility of experimental realization of spin polarized transport is more favorable via edge-fluorination than that of the edge hydrogenation. Moreover, passivation of only edge-B (N) atoms with F atoms i.e. ABNNR$_{FB}$ (ABNNR$_{FN}$) transforms insulating bare ABNNRs into half-metal (wide band gap semiconductor) and gives rise to a substantial magnetic moment of about 2.03 (2.02) µ$_B$. Conversely, both-edge F-passivation (ABNNR$_{FBN}$) does not alter electronic or magnetic behavior significantly; therefore, ABNNR$_{FBN}$ structures are insulators with degenerate/symmetric spin states. Owing to the possibility of promising potential of spin polarization, we expect experimental realization of half-metallicity in ABNNRs via edge fluorination.

## ACKNOWLEDGMENTS

The authors are thankful to Computational Nanoscience and Technology Labarotary (CNTL), ABV—Indian Institute of Information Technology and Management Gwalior (India) for providing computational facility. One of the authors (HMR) acknowledges the Ministry of Human Resource Development (MHRD), government of India for providing financial support as Teaching Assistantship.

## REFERENCEES:

# **Tables:**

Table 1. Calculated cohesive energies (eV) as a function of ribbon width for all considered edge fluorinated ABNNR structures.

| Ribbon width $N_a$ | $E_c$ (eV) | | |
|---|---|---|---|
| | ABNNR$_{FB}$ | ABNNR$_{FBN}$ | ABNNR$_{FN}$ |
| 6 | -9.08 | -8.54 | -8.56 |
| 7 | -9.22 | -8.73 | -8.77 |
| 8 | -9.33 | -8.87 | -8.93 |
| 9 | -9.41 | -9.00 | -9.06 |
| 10 | -9.49 | -9.10 | -9.16 |

Table 2. Calculated binding energy (eV) per F (or H)-atom as a function of ribbon width for all considered F and H-passivated ABNNR structures; the values in brackets are corresponding to H-passivation.

| Ribbon width $N_a$ | $E_B$ (eV) | | |
|---|---|---|---|
| | ABNNR$_{FB(HB)}$ | ABNNR$_{FBN(HBN)}$ | ABNNR$_{FN(HN)}$ |
| 6 | -6.406 (-3.489) | -5.588 (-5.184) | -2.792 (-4.178) |
| 7 | -6.424 (-3.493) | -5.594 (-5.188) | -2.804 (-4.189) |
| 8 | -6.416 (-3.486) | -5.591 (-5.186) | -2.797 (-4.184) |
| 9 | -6418 (-3.482) | -5.592 (-5.187) | -2.800 (-4.186) |
| 10 | -6.417 (-3.482) | -5.592 (-5.187) | -2.799 (-4.185) |

Table 3. Calculated magnetic moment and energy difference ($\Delta E=|E_{LSDA}-E_{LDA}|$) between spin polarized and spin un-polarized states for most stable ABNNR$_{FB}$ structures with varying ribbon width.

| | $N_a$=6 | $N_a$=7 | $N_a$=8 | $N_a$=9 | $N_a$=10 |
|---|---|---|---|---|---|
| $M$ ($\mu_B$) | 2.03 | 2.04 | 2.03 | 2.03 | 2.03 |
| $\Delta E$ (eV) | 0.41 | 0.38 | 0.40 | 0.42 | 0.41 |

Table 4. The overall magnetic and electronic behavior for all considered edge fluorinated ABBNRs ($N_a$=6 to 10).

| Ribbon Configuration | Avg. $M$ ($\mu_B$) | Electronic state | Band gap (eV) | Magnetic behavior |
|---|---|---|---|---|
| ABNNR$_{FB}$ | 2.03 | Half-metal (0.21 eV) | 0 | Magnetic |
| ABNNR$_{FBN}$ | 0 | Insulator | 4.26 to 4.38 | Non- magnetic |
| ABNNR$_{FN}$ | 2.02 | Wide band gap Semiconductor | 3.16 to 3.34 | Magnetic |

**Figures:-**

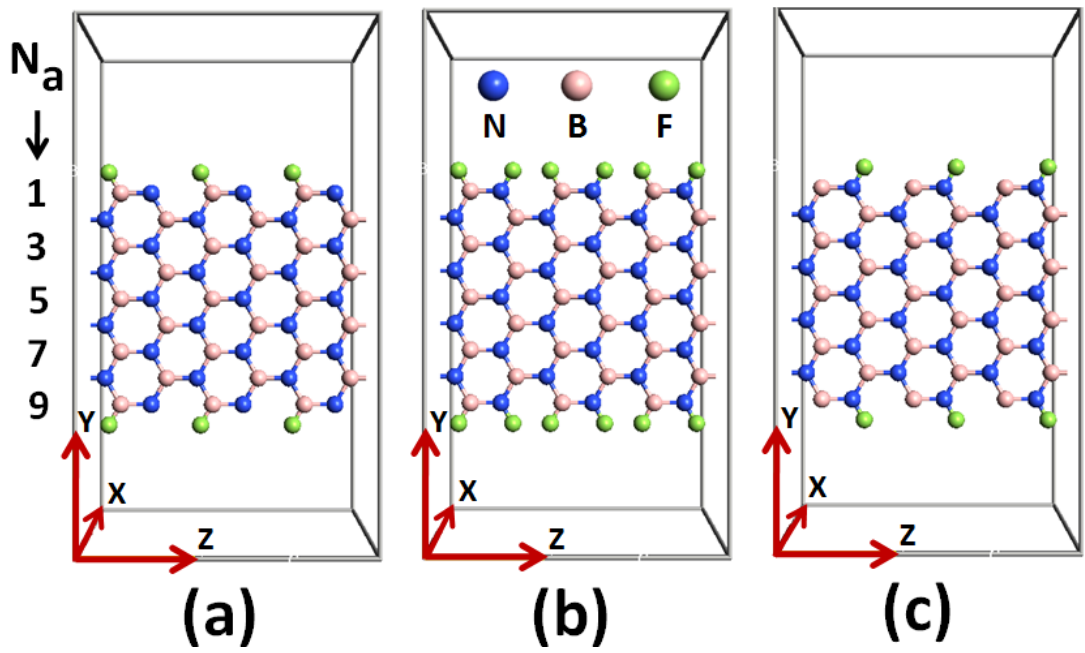

**Figure 1     Rai et al.**

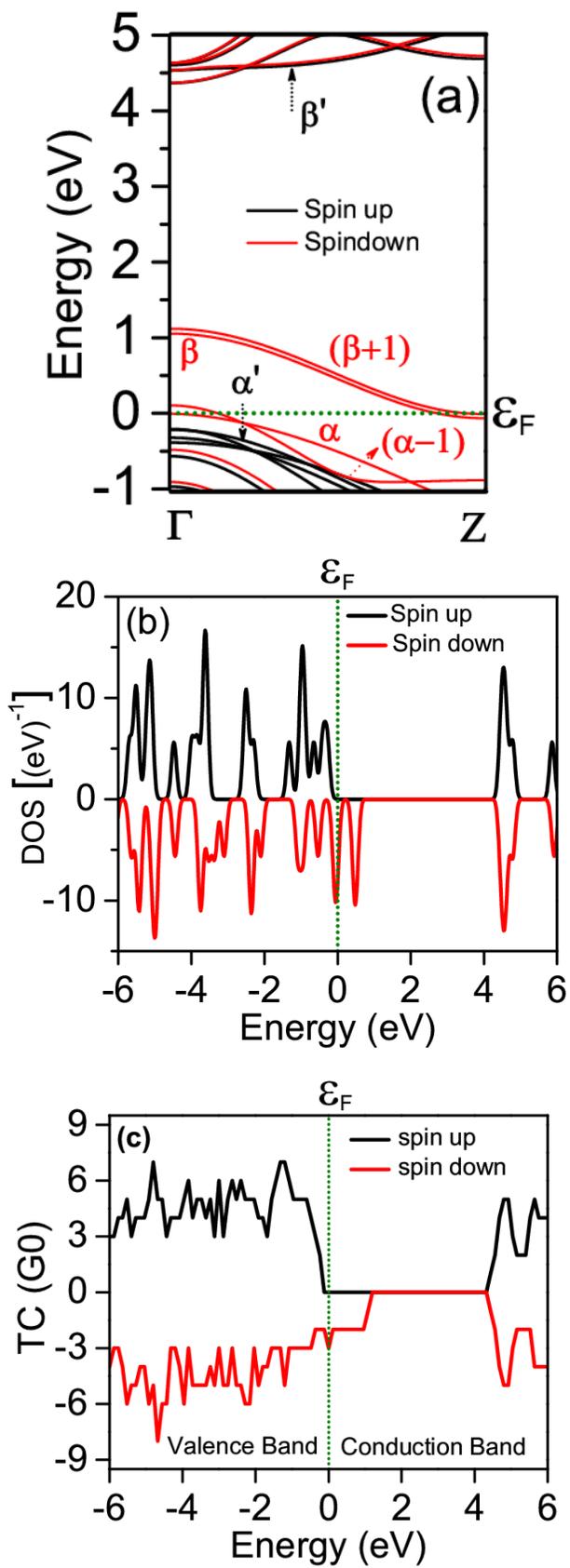

Figure 2    Rai et al.

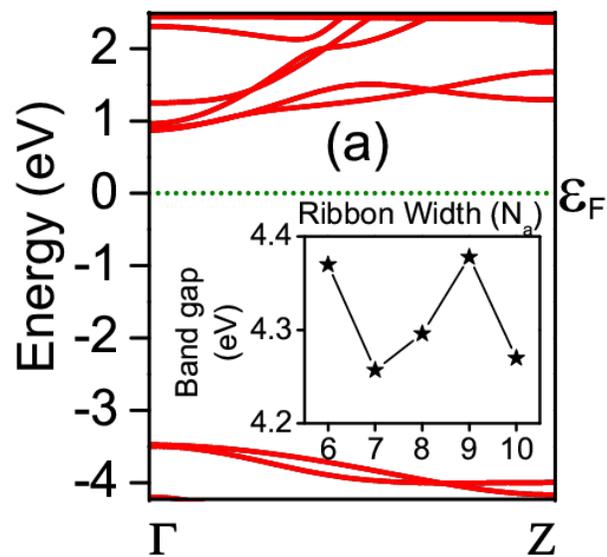

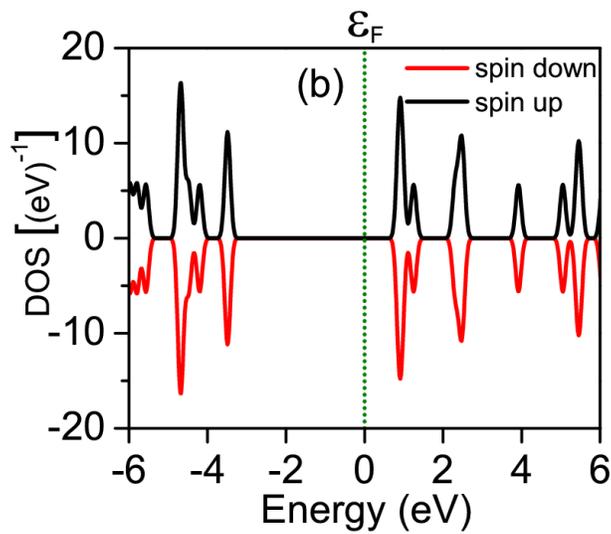

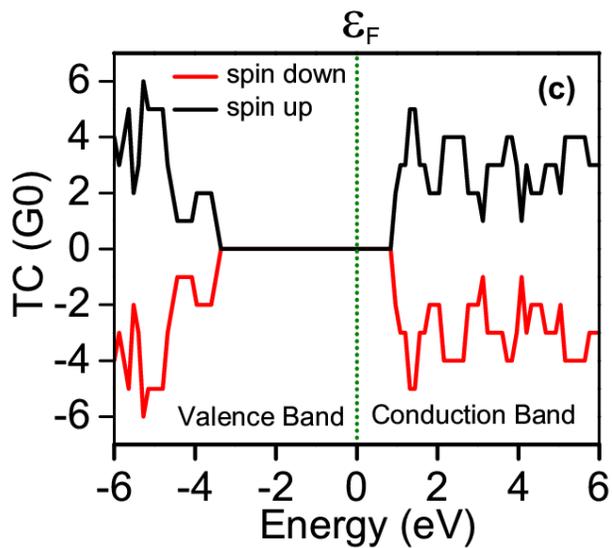

**Figure 3    Rai et al.**

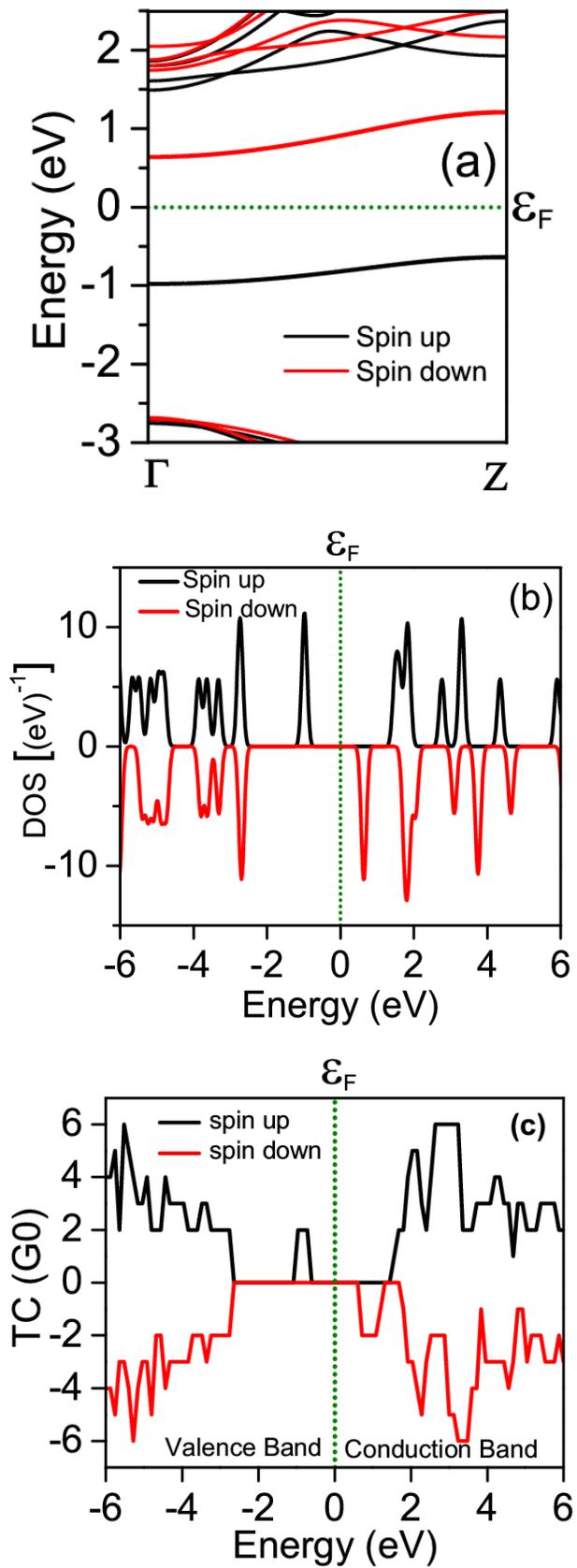

Figure 4      Rai et. al

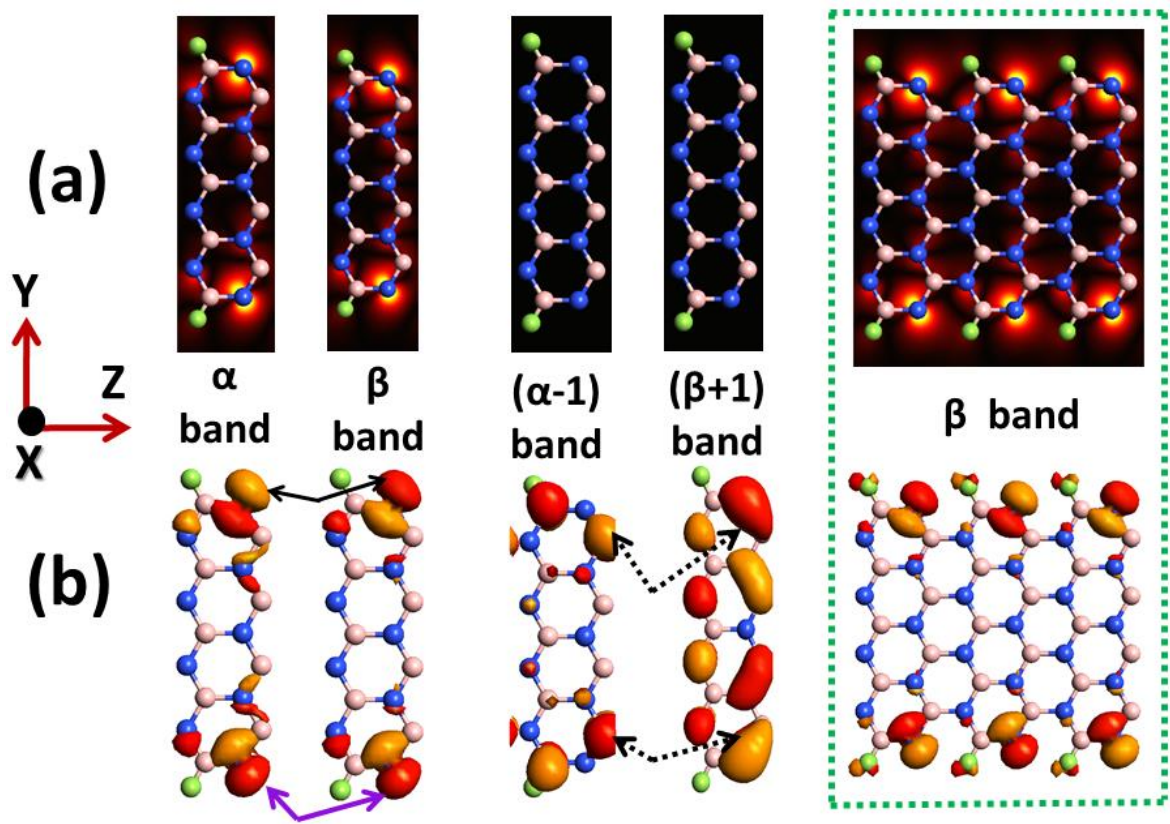

Figure 5   Rai et. al.

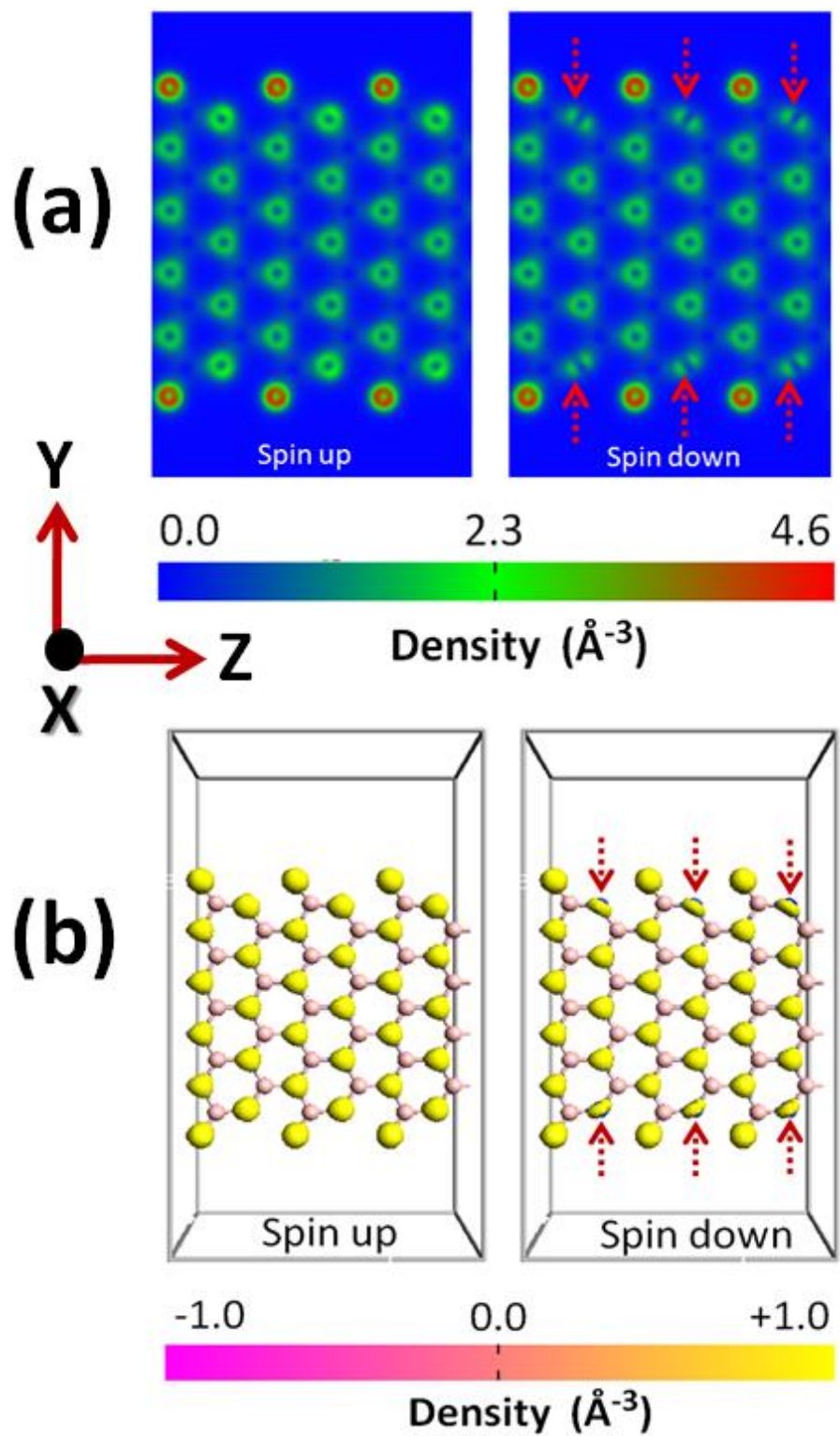

**Figure 6    Rai et. al.**

**Figure Captions:**

FIG. 1 (Color online). Convention of ribbon width and the supercell units ($N_a$=9 as a representative) for (a) ABNNR$_{FB}$, (b) ABNNR$_{FBN}$ and (c) ABNNR$_{FN}$. The ABNNRs are modeled with periodic boundary conditions along *z*- axis whereas x and y directions are confined.

FIG. 2 (Color online). Half-metallic behavior of 9-ABNNR$_{FB}$; calculated (a) electronic band structure and (b) density of states (DOS) and (c) transmission spectrum (TS) for 9-ABNNR$_{FB}$. In (a), spin down (spin-up) states nearest to the Fermi level in valence band and conduction band are indexed with α (α') and β (β') respectively. The Fermi level is indicated by green dotted line whereas spin down (spin-up) states are shown with red (black) color.

FIG. 3 (Color online). Spin degeneracy and insulating nature of 9-ABNNR$_{FBN}$; Calculated (a) electronic band structure and (b) density of states (DOS) and (c) transmission spectrum (TS) for 9-ABNNR$_{FBN}$. The Fermi level is indicated by green dotted line whereas spin down (spin-up) states are shown with red (black) color.

FIG. 4 (Color online). Calculated (a) electronic band structure and (b) density of states (DOS) and (c) transmission spectrum (TS) for 9-ABNNR$_{FN}$. The Fermi level is indicated by green dotted line whereas spin down (spin-up) states are shown with red (black) color.

FIG. 5 (Color online). Calculated Bloch states for spin down α, β, (α-1), (β+1) bands and spin up α'and β' bands of 9-ABNNR$_{FB}$ are presented in the form of (a) Contour and (b) Isosurface. The isovalue is 0.09 (au). To keep the size of figure in a presentable format only one third (along z-axis) of actual unit cell is shown and for a clear view, the results corresponding to β band are presented also with actual ribbon geometry (enclosed in a dotted rectangle) which is used for calculations. The ABNNRs are modeled with periodic boundary conditions along *z*- axis whereas x and y directions are confined. B, N and F atoms are same in shape, size and color as that of represented in figure1.

FIG. 6 (Color online). The real space distribution of calculated electron densities corresponding to spin up and spin down electrons for 9-ABNNR$_{FB}$ is presented in the form of (a) Contour and (b) Isosurface. The corresponding density scale is shown by a tricolored rectangular horizontal strip situated at the bottom. In figure (b), the electron density is plotted corresponding to isovalue 1 (au). In order to display the distinct change between spin up and spin down electron densities at ribbon edges, ribbon configurations are not shown in figure (a). B, N and F atoms are same in shape, size and color as that of represented in figure1. The ABNNRs are modeled with periodic boundary conditions along *z*- axis whereas x and y directions are confined.

# Possibility of Spin-polarized Transport in Edge Fluorinated Armchair Boron Nitride Nanoribbons

Hari Mohan Rai[*1], Shailendra K. Saxena[1], Vikash Mishra[1], Ravikiran Late[1], Rajesh Kumar[1], Pankaj R. Sagdeo[1], Neeraj K. Jaiswal[2] and Pankaj Srivastava[3]

[1]Material Research Lab. (MRL), Indian Institute of Technology Indore, Simrol, Indore (M.P.) – 452020, India
[2]PDPM- Indian Institute of Information Technology, Design and Manufacturing, Jabalpur – 482005, India
[3]Computational Nanoscience and Technology Lab. (CNTL),
ABV- Indian Institute of Information Technology and Management, Gwalior – 474015, India
* Corresponding author: pankajs@iiitm.ac.in

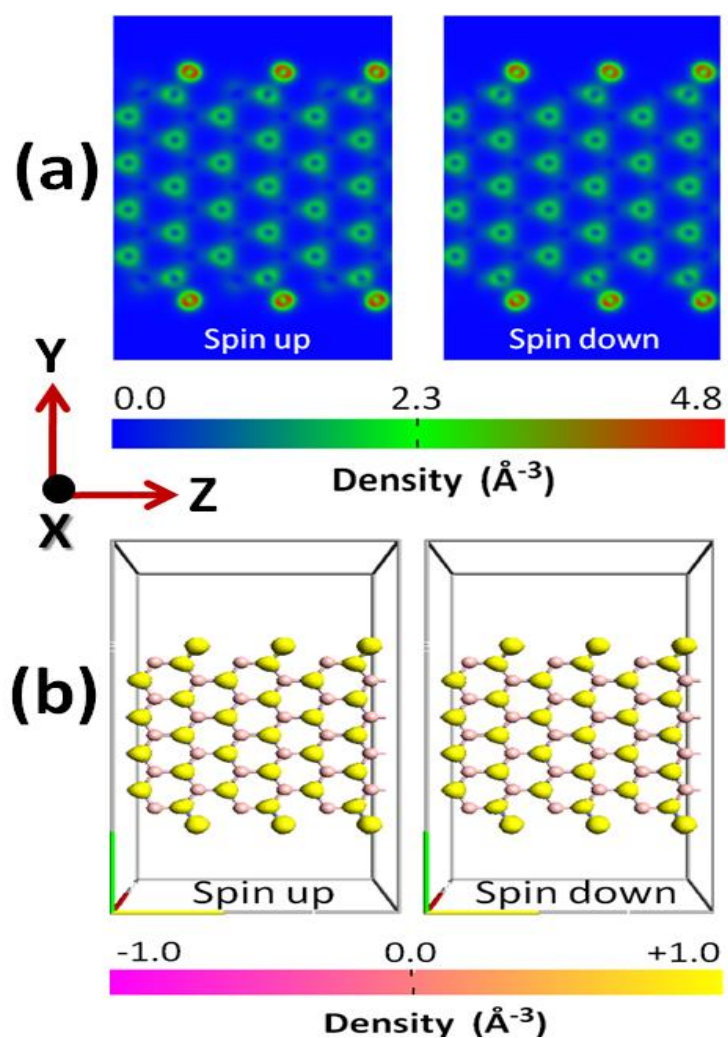

**Fig. S1:** (Color online). The real space distribution of calculated electron densities corresponding to spin up and spin down electrons for 9-ABNNR$_{FN}$ is presented in the form of (a) Contour and (b) Isosurface. The corresponding density scale is shown by a tricolored rectangular horizontal

strip situated at the bottom. In figure (b), the electron density is plotted corresponding to isovalue 1 (au). In order to display the distinct change (if any) between spin up and spin down electron densities at ribbon edges, ribbon configurations are not shown in figure (a). B, N and F atoms are same in shape, size and color as that of represented in figure1. The ABNNRs are modeled with periodic boundary conditions along *z*- axis whereas x and y directions are confined.

We found that the fluorination of the zigzag ribbons is energetically more favorable than that of the ABNNRs. The calculated values for the ribbons with fluorinated zigzag edges are given below for comparison with edge-fluorinated ABNNRs (tabulated in main text).

**Table S1:** Calculated **cohesive energy Ec** (eV) as a function of ribbon width for fluorinated ZBNNR structures.

| Ribbon width $N_Z$ | $E_c$(eV) | | |
|---|---|---|---|
| | $ZBNNR_{FB}$ | $ZBNNR_{FBN}$ | $ZBNNR_{FN}$ |
| 6 | -9.588 | -9.305 | -9.283 |
| 8 | -9.734 | -9.506 | -9.501 |
| 10 | -9.825 | -9.634 | -9.636 |

**Table S2:** Calculated **binding energy $E_B$** (eV) per F-atom as a function of ribbon width for fluorinated ZBNNR structures.

| Ribbon width $N_Z$ | $E_B$(eV) | | |
|---|---|---|---|
| | $ZBNNR_{FB}$ | $ZBNNR_{FBN}$ | $ZBNNR_{FN}$ |
| 6 | -9.450 | -7.535 | -5.483 |
| 8 | -9.479 | -7.551 | -5.511 |
| 10 | -9495 | -7.558 | -5.527 |